\documentclass{aa}
\usepackage{graphicx}
\newcommand{\ergcm}{\mbox{ erg cm$^{-2}$}}
\newcommand{\ergcms}{\mbox{erg cm$^{-2}$ s$^{-1}$}}
\def\sax{{\it BeppoSAX}}
\begin{document}

\title{The dark burst 010214 with \sax: possible variable absorption and jet X--ray emission}

\author{C. Guidorzi\inst{1},
F. Frontera\inst{1,2},
E. Montanari\inst{1,7},
L.~Amati\inst{2},
L.A. Antonelli\inst{3},
J.J.M.~in~'t~Zand\inst{4}, 
E.~Costa\inst{5},
R.~Farinelli\inst{1},
M.~Feroci\inst{5},
J.~Heise\inst{4},
N.~Masetti\inst{2},
L.~Nicastro\inst{6},
M.~Orlandini\inst{2},
E.~Palazzi\inst{2},
L.~Piro\inst{5}
}

\offprints{C. Guidorzi,\\\email{guidorzi@fe.infn.it}}

\institute{$^1$Dipartimento di Fisica, Universit\`a di Ferrara,
via Paradiso 12, I-44100 Ferrara, Italy\\
$^2$Istituto Astrofisica Spaziale e Fisica Cosmica, Sezione di Bologna, CNR,
via Gobetti 101, I-40129 Bologna, Italy\\
$^3$Osservatorio Astronomico di Roma,
via Frascati 33, Monteporzio I-00040, Italy\\
$^4$Space Research Organization in the Netherlands,
 Sorbonnelaan 2, 3584 CA Utrecht, The Netherlands\\
$^5$Istituto Astrofisica Spaziale e Fisica Cosmica, Sezione di Roma, CNR, 
via Fosso del Cavaliere, I-00133 Roma, Italy\\
$^6$Istituto Astrofisica Spaziale e Fisica Cosmica, Sezione di Palermo, CNR,
via U. La Malfa 153, I-90146 Palermo, Italy\\
$^7$ITA ``I. Calvi'', Finale Emilia (MO), Italy
}
\date{Received \date; Accepted \dots}

\abstract{
We report on the prompt and afterglow emission observations of the dark
burst GRB010214  with \sax. The prompt emission shows possible evidence 
of variable absorption from $N_{\rm H} = 3.0^{+5.1}_{-2.0} \times 
10^{23}$~cm$^{-2}$ in the first 6 s of the event to a value consistent 
with the Galactic column density
($N_{\rm H}^{\rm G} = 2.66\times 10^{20}$~cm$^{-2}$) in the GRB
direction. 
An X--ray afterglow emission in the 2--10 keV energy band was detected 
with \sax, but an analogue search at lower wavelengths (optical, IR and 
radio) was unsuccessful.  
The X--ray afterglow spectrum is consistent with a power--law with 
Galactic absorption. The light curve shows a complex decay, if
the tail of the prompt emission is assumed as the onset of the afterglow:
if the origin of the afterglow is coincident with the GRB onset,
a bump before $\sim 3\times10^4$~s is inferred, while if the afterglow 
is assumed to start later, a steepening of the power--law light curve 
at  $t\sim 3\times10^4$~s is deduced. We discuss these results in the light
of the current models of afterglows and the possible origin of the
GRB darkness. Finally, we tentatively derive an estimate of the
burst redshift.

\keywords{gamma rays: bursts --- gamma rays: observations}}

\maketitle
\markboth{C. Guidorzi et~al.: GRB010214 with \sax}{}

\section{Introduction}
\label{s:intro}
The available sample of more than 30 
X--ray afterglows of Gamma-Ray Bursts (GRBs) detected with \sax\ 
(\cite{Frontera02,Piro02}) shows a variety of emission
properties that, even if they can be basically accounted for in the context of the
fireball model (see, e.g., review
by Piran 1999, 2000),\nocite{Piran99,Piran00} shed light on several details
of the GRB phenomenon. For example, breaks in the X-ray light curves, inferred
by comparing the late afterglow emission with the prompt
emission from GRB990510 (\cite{Pian01}) and GRB010222 (\cite{Zand01}),
not only confirm the breaks observed in the optical afterglow emission  
of these GRBs (Stanek et~al. 1999, \nocite{Stanek99} and Harrison
et~al. 1999,\nocite{Harrison99} for GRB990510; Masetti et~al. 2001a,
\nocite{Masetti01a} for GRB010222) but also give support to the fact that
the onset of the X--ray afterglows occurs during the tail of the prompt
GRB emission (\cite{Frontera00}). Achromatic breaks in the afterglow
emission can be an imprint of a relativistic jet when it slows down and
spreads laterally (\cite{Sari99}), but can also be, e.g., the consequence of a transition
of the relativistic fireball to a non-relativistic regime (NR) in a dense
circumburst medium (\cite{Dai99}). The different relationships expected between
the slope of the afterglow light curve and that of the energy spectrum
can help to distinguish between the various models and allow to gain information
on the circumburst medium properties. In
the case of GRB010222, In~'t Zand et~al. (2001) and Masetti et~al. (2001a)
find that the NR effect is more consistent with the data.
In this respect, some authors suggest that jet effects could not be the unique
reason for such breaks (\cite{Wei02}).
The continuum energy spectra of the X--ray afterglows are generally well
fit with power--laws. However their interpretation is not unique: they can
be either due to synchrotron emission, as in the case of GRB970508 (\cite{Galama98})
or to inverse--Compton emission as in the case
of GRB000926 (\cite{Harrison01}). Only broad band spectra, that cover frequency
ranges from the radio to X--rays, can provide the handle on the right model.
Detection of inverse-Compton emission is a strong hint for the presence of a
dense medium around the GRB location.

X--ray afterglows have been measured in 86\% of the localized GRBs 
(e.g., Frontera 2002, Piro 2002),\nocite{Frontera02,Piro02} but only 50\% of the GRBs
with X--ray afterglows show optical (OTs) and/or radio transients (RTs).
The explanation for the non--detection of OTs is not well established.
A possible interpretation of such "dark bursts" in visible light is that
they are obscured by dust in their host galaxies (see reviews by Djorgovski 2001
and Pian 2002\nocite{Djorgovski01,Pian02}). This interpretation is likely to be valid
when we observe radio transients (RTs) without OTs; in the remaining cases the issue 
is still open.
A possibility is that dark bursts are at very high redshifts ($z> 4.5$), as
first suggested by Fruchter (1999).\nocite{Fruchter99}
The shape of the light curve of the X-ray afterglows can provide a 
hint on the presence of dense star--forming regions around GRBs:
a bump followed by a steeper decay in soft X--rays is predicted for bursts
that are heavily obscured in the optical (\cite{Meszaros00}).

In these respects GRB010214 shows interesting features which we will report and
discuss in this paper. As we will see, it showed an X--ray afterglow emission
but no positive signal was detected either in the optical or in the radio bands.
Thus, it belongs to the class of ``dark'' bursts. 

\section{Observations and data analysis}
\label{s:obs}
GRB010214 was detected on February 14, 2001, at 08:48:11 UT with the
\sax\ (\cite{Boella_97}) Gamma-Ray Burst Monitor
(GRBM, 40--700~keV, Frontera et~al. 1997\nocite{Frontera97}), 
and the Wide Field Camera No. 2 (WFC, 2--28~keV, Jager et~al.\ 1997).\nocite{Jager97}
The GRB position was promptly estimated with an error
circle of radius 3$'$ (\cite{Gandolfi_gcn933}).
The BeppoSAX Narrow Field Instruments (NFI) started observing
the GRB error box 6.28 hrs after the burst; the observation lasted 
1.91 days, with a net exposure time of 83 ks for the MECS and 23 ks for the LECS.
A new uncatalogued X--ray source was detected within the WFC error
circle in both MECS and LECS
at coordinates $\alpha_{2000.0} = 17^{\rm h}40^{\rm m}58^{\rm s},
\delta_{2000.0} = +48\degr34\arcmin37\arcsec$ with an
error radius of 1$'$ (\cite{Frontera_gcn950,Guidorzi_gcn951}), showing an apparent 
decaying flux. Optical/IR/radio counterpart observations of the GRB
error box did not find any fading source (see Klose et~al.2001a, Zhu \& Xue 2001,
Hudec et~al. 2001, Berger \& Frail 2001, Antonelli et~al. 2001,
Uemura et~al. 2001, Gorosabel et~al. 2001a and 2001b, Masetti et~al. 2001b,
Rol et~al. 2001a, Cowsik 2001, Henden 2001), in spite of some initial IR
(\cite{Dipaola01}) and optical (\cite{Rol01b,Klose01b}) claim of afterglow candidate.
Among the upper limits reported above, the most constraining ones are shown
in fig.~\ref{f:R-upp-lim}.

Data available from GRBM  include two 1~s ratemeters in two energy
channels (40--700~keV and $>$100~keV), 128~s  count spectra (40--700~keV,
225 channels) and high time resolution data (down to 0.5~ms) in the
40--700~keV energy band.
WFCs were operated in normal mode with 31 channels in the 2--28~keV energy band and 
0.5~ms time resolution (\cite{Jager97}). 
The burst direction was offset by 7$^\circ$ with respect to
the WFC axis. With this offset, the
effective area exposed to the GRB was
$\approx$~510~cm$^2$ in the 40-700~keV band and  72~cm$^2$
in the 2--28~keV energy band.
The background  in the WFC and GRBM energy bands  was fairly stable during the event.
The GRBM background level was estimated by linear
interpolation using the 250~s count rate data  before and after the burst.
The WFC spectra were extracted through the Iterative Removal Of Sources
procedure (IROS, e.g. Jager et~al. 1997)\nocite{Jager97} which implicitly subtracts the
contribution of the background and of other point sources in the field of
view.

%
\begin{figure}[!h]
\centerline{\includegraphics[width=8.5cm,height=8.5cm]{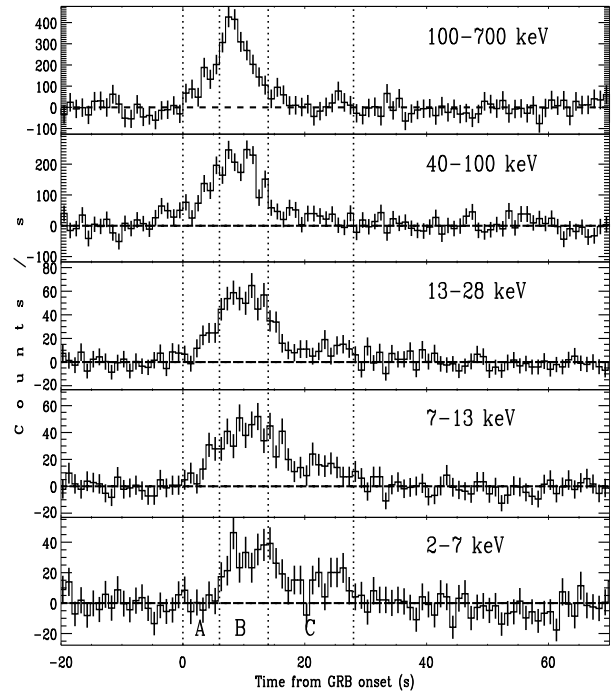}}
\vspace{0.5cm}
\caption[]{
Light curves of GRB010214 in five energy bands,
after background subtraction. The zero abscissa corresponds to 2001 February 14, 
08:48:05.1 UT. The three time intervals A, B and C 
over which the spectral analysis has been performed are indicated
with vertical dotted lines. 
}
\label{f:lc}
\end{figure}

We performed a joint spectral analysis of the LECS and MECS 2+3 data. The source
was apparent in the 2--10 keV energy band, but did not show any statistically
significant signal in the 0.1--2.0 keV band covered by the LECS.
In 2--10 keV the source was apparent up to 72 ks from the GRB onset. 
In the last part of the source observation, from 72 to 187 ks, the source was no 
more visible.

The spectral fitting was performed using the XSPEC package v. 11.0.1
(\cite{Arnaud96}). The elemental abundances are with respect to those by Anders \& 
Ebihara (1982)\nocite{Anders82} and the opacities are from Morrison \& McCammon 
(1983)\nocite{Morrison83}. The quoted errors are intended to be at 90\% confidence
level (cl) for one parameter ($\Delta \chi^2 = 2.7$), except when otherwise
specified.


\section{Results} 
\label{s:results}
\subsection{Prompt Emission}
\label{s:prompt}

The GRBM and WFC light curves in different energy bands are shown
in Fig.~\ref{f:lc}. Above 7 keV, the GRB shows a single pulse, which broadens with
decreasing energy. A tail after the peak in the WFC energy band 
is also apparent. In the lowest energy band (2--7 keV), the main pulse shortens
and the tail becomes a secondary peak. The  total duration of the burst in the GRBM
passband is $\sim$15 ~s, while in the low energy X--ray band it is
$\sim$30~s.

The time-averaged 2--700 keV GRB spectrum was derived using the WFC data and
the 225 channel GRBM count data.
A simple power--law ({\sc pl}) does not fit ($\chi^2/dof = 38.5/17$) the data, which 
are instead well fit ($\chi^2/dof = 11.7/16$) 
with a photoelectrically absorbed broken power--law ({\sc bknpl})  with photon 
index $\Gamma_X = 0.86^{+0.13}_{-0.17}$ below the break energy $E_0 = 67^{+20}_{-15}$ keV 
and $\Gamma_\gamma = 2.4^{+0.3}_{-0.2}$ above $E_0$ (see Table~\ref{t:tab1}). 
The fit with a smoothly broken 
power--law ({\sc bl}, Band et~al. 1993\nocite{Band93}) does not allow the 
determination of the high energy photon index. By fixing it to the value derived from 
the {\sc bknpl} model, the fit results
are reported in Table~\ref{t:tab1}. The $N_{\rm H}$ is consistent with
the Galactic value $N_{\rm H}^{\rm G} = 2.66\times 10^{20}$~cm$^{-2}$ along the
GRB direction ($l = 75.3^{\circ}$, $b = +31.3^{\circ}$). 
From the time-averaged GRB spectrum, we derived 
a GRB 1~s peak flux of $(7.8 \pm 1.3)\times 10^{-7}\ergcms\ $ and a 40--700 keV fluence
$S_{40-700} = (4.5 \pm 0.8)\times 10^{-6}\ergcm\ $. In the 2--28 keV band the fluence is 
$S_{2-28} = (7.3 \pm 1.1)\times 10^{-7}\ergcm\ $, with a $S_{2-28}/S_{40-700}$ ratio of
about 0.16. For comparison with the BATSE GRBs, the 50--300 keV fluence of GRB010214
is $S_{50-300} = (3.3 \pm 0.7)\times 10^{-6}\ergcm\ $ with a hardness ratio 
$C(100-300)/C(50-100) = 1.2 \pm 0.3$. 

%
\begin{table*}
\label{table1}
\begin{center}
\caption{Best-fit parameters of the GRB010214 time-averaged
spectrum and of its temporal evolution (see text for the definition of
slices A, B and C). Errors are at 90\% confidence level.}
\begin{tabular}{cccccccc}
\hline\hline
Slice & Duration (s) & Model($^{\mathrm a}$) & $N_{\rm H}$($^{\mathrm b}$)
& $\Gamma_X$ & $\Gamma_\gamma$ & E$_{p}$~(keV)($^{\mathrm c}$) & $\chi^{2}/\nu$\\
\hline
A+B+C  & 28 & {\sc bknpl} & [0.0266] & $0.86^{+0.13}_{-0.17}$ & $2.4^{+0.3}_{-0.2}$ & 
            $67^{+20}_{-15}$ & 11.7/16\\
       &    & {\sc bl} & [0.0266] & $0.65^{+0.16}_{-0.19}$ &  [2.4] & $102^{+42}_{-29}$ & 
            14.0/17\\ \\      
 A     & 6  & {\sc pl} & $30^{+51}_{-20}$ & $1.6\pm 0.2$  & -- & $>700$ & 4.66/4  \\
       &    & {\sc bknpl} & [0.0266] & $0.34^{+0.43}_{-0.50}$ & $2.5^{+0.8}_{-0.9}$
            &  $41^{+39}_{-25}$ & 0.86/3 \\ \\
 B     & 8  & {\sc bknpl} & [0.0266] & $0.91^{+0.15}_{-0.15}$ & $>1.8$ & $75^{+105}_{-37}$ 
            & 1.8/7 \\
       &    & {\sc bl} & [0.0266] & $0.73^{+0.24}_{-0.27}$ & $>1.7$ & $122^{+113}_{-46}$ 
            & 2.6/7 \\ \\
 C     & 14 & {\sc bknpl} & [0.0266] & [0.81]  & $>1.9$ & $10^{+9}_{-6}$ 
            & 3.8/2 \\
\hline
\end{tabular}
\begin{list}{}{}
\item ($^{\mathrm a}$){\sc pl} = power--law; {\sc bknpl} = broken power--law; 
{\sc bl} = smoothly broken power--law (\cite{Band93});
\item ($^{\mathrm b}$)units of $10^{22}$~cm$^{-2}$;
\item ($^{\mathrm c}$)peak energy of the $EF(E)$ spectrum.
\end{list}
\label{t:tab1}
\end{center}
\end{table*}

\begin{figure}[!h]
\centerline{\includegraphics[width=8.5cm]{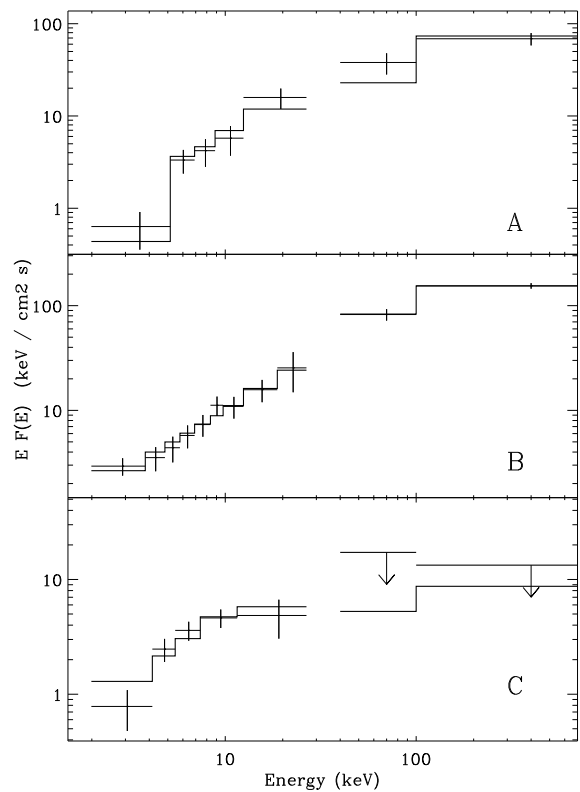}}
\vspace{0.6cm}
\caption[]{
$E F(E)$ spectrum of the prompt emission in the A, B, and C intervals. The shown upper 
limits are at 2$\sigma$ level. For the C interval, the {\sc bknpl} model assuming
$\Gamma_\gamma = 2.1$ is shown (see Table~\ref{t:tab1}).
}
\label{f:nuFnu}
\end{figure}

The results of the investigation on the spectral evolution of the GRB prompt emission 
are reported in Table~\ref{t:tab1} and Fig.~\ref{f:nuFnu}.
The GRB light curve was subdivided in three time intervals A, B and C (see Fig.~\ref{f:lc})
and the spectrum of each interval derived. The spectrum of the interval A
is found to be well fit by a photoelectrically absorbed {\sc pl} model, with column 
density $N_{\rm H} = 3.0^{+5.1}_{-2.0} \times 10^{23}$~cm$^{-2}$,  significantly in excess
of the Galactic $N_{\rm H}^{\rm G}$. Freezing $N_{\rm H}$ to the Galactic value 
($N_{\rm H}^{\rm G}$), the fit is not acceptable ($\chi^2/dof = 15.8/5$). An alternative
to the {\sc pl} model is the {\sc bknpl} model, which fit the data (see Table~\ref{t:tab1})
even assuming a Galactic absorption. However the peak energy $E_p$ derived with this model 
($\sim 40$~keV) contradicts the usual hard-to-soft spectral evolution of GRBs. For this
reason, we prefer the {\sc pl} model, which implies an initial $E_p > 700$~keV.

Unlike the A spectrum, the B spectrum cannot be fit with a {\sc pl} model with either 
Galactic ($\chi^2/dof = 18/9$) or higher $N_{\rm H}$ ($\chi^2/dof = 15.1/8$).
It can  be fit with either a {\sc bknpl} or a {\sc bl} model with 
Galactic column density (see Table~\ref{t:tab1}). Finally
the C spectrum is mainly constrained by the WFC data, given that we have only
upper limits with GRBM. It can be fit by a {\sc bknpl} model 
with Galactic absorption. Freezing the low energy photon index to the value
(0.81) found from the fit with a {\sc pl} with an exponential cutoff, 
the other parameter values  are reported in Table~\ref{t:tab1}.

%
%

\begin{figure}[!h]
\centerline{\includegraphics[width=8.0cm]{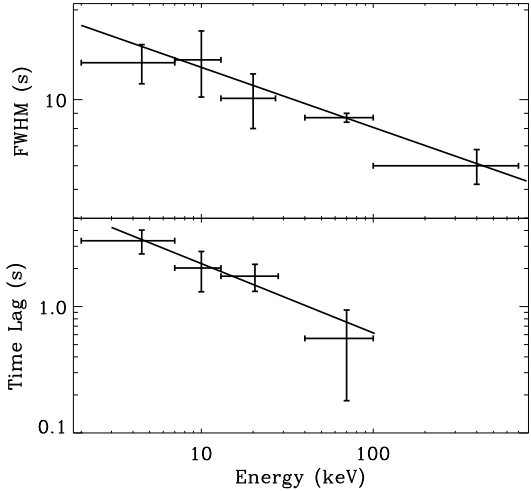}}
\vspace{0.6cm}
\caption[]{{\it Top panel:} pulse FWHM as a function of energy with superposed
the best power--law fit. {\it Bottom panel:} lag of the pulse centroid with respect
to that in the 100--700 keV band as a function of energy, with superposed the power--law 
best fit.}
\label{f:FWHM_lag}
\end{figure}

From the visual inspection of the 1 s light curves (Fig.~\ref{f:lc})
it is apparent that the pulse peak in the higher energy band leads that at 
lower energies while the pulse width decreases. This property is not new 
(\cite{Fenimore95,Norris96}). We have evaluated, as a function of energy,
both  the time lag $\Delta T$ of the pulse peak with 
respect to the 100--700 keV peak time and the pulse FWHM  by fitting the 
pulse time profiles shown in Fig.~\ref{f:lc} with a Gaussian plus a polynomial. 
The results are shown in Fig.~\ref{f:FWHM_lag}. Both FWHM and $\Delta T$ vs. energy
are well fit with a power--law ($\propto E^{-\alpha}$). The  best fit
index $\alpha_{\rm FWHM}$  is either  $0.20 \pm 0.10$ or $0.28 \pm 0.13$, depending
whether the midpoints or the lower bounds of the energy bands are taken into account,
respectively, while the best fit power--law index $\alpha_{\Delta T}$ 
is $0.55 \pm 0.17$.
The width of the main pulse in the energy channel below 7~keV (see Fig.~\ref{f:lc})
shortens instead of broadening, but a secondary pulse rises. Only the sum of their
FWHM, evaluated by rebinning the 2--7~keV light curve at 5~s, is consistent with
the power--law slope evaluated at higher energies (see Fig.~\ref{f:FWHM_lag}).
%
%

\subsection{Afterglow Emission}
\label{s:afterglow}

\begin{figure}
\centerline{\includegraphics[width=7.5cm]{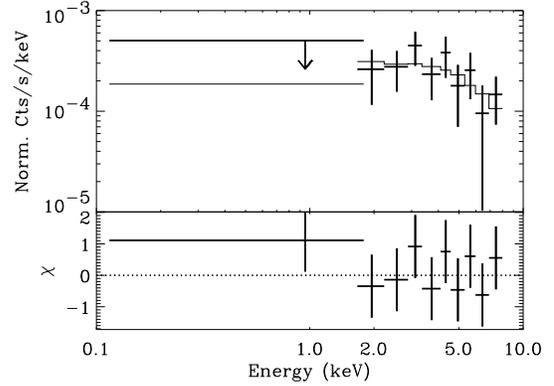}}
\caption[]{
LECS+MECS23 spectrum of the X--ray afterglow.
The best power--law ($\Gamma = 1.3$) fit is shown. The 2-$\sigma$ upper 
limit comes from the LECS. In the bottom panel, the residual in
correspondence of the LECS upper limit is the measured
value in the LECS.
}
\label{f:aft_spectrum}
\end{figure}

The entire observation of 165 ks duration was split into two parts, one of 49.4 ks
duration during which the source was visible and fading, and the other of 115.6 ks
duration, during which the source was no more visible.
The afterglow spectrum of the first part is shown in Fig.~\ref{f:aft_spectrum}. Given that
below 1 keV the source was detected only marginally,  the 2$\sigma$ upper limit is shown.
The spectrum is well fit with a photoelectrically absorbed power--law with  
Galactic column density $N_{\rm H}^{\rm G}$ and photon index
$\Gamma = 1.3^{+0.8}_{-0.6}$. No evidence of spectral evolution with time 
was noticed. Assuming the best fit spectrum, we derived a 3$\sigma$ upper limit
of $9\times 10^{-14}\ergcms\ $ to the 2--10 keV source flux during the second
part of the observation (starting at 72 ks after the main event).

%
%

\begin{figure}
\centerline{\includegraphics[width=8.0cm]{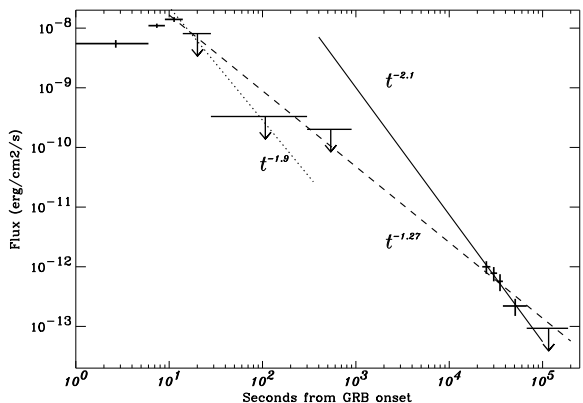}}
\caption[]{
2--10 keV flux decay from the prompt emission, assuming
as origin the GRB onset.
The dotted line indicates the minimum power--law decay slope on the basis of
the WFC upper limits; the dashed line  is the best-fit power--law decay
of the last significant WFC data point and the NFI data; the solid line shows the
best fit power--law  when only the afterglow measurements are
taken into account.
}
\label{f:aft_lc0}
\end{figure}

Figure~\ref{f:aft_lc0} shows the 2--10 keV flux decay from the prompt emission (WFC data)
to the afterglow (MECS2+3), where the time origin is the GRB onset. Using a power--law 
model for the decay law ($F_X \propto t^{-\delta}$),
the NFI afterglow measurements are well fit with an index $\delta = 
2.1^{+1.0}_{-0.6}$. If we include in the fit the last
significant point of the prompt emission, we obtain an index $\delta = 1.27 \pm 0.04$, 
which however is not consistent with the $3\sigma$
upper limit provided by the WFC from 30~s to 300~s after the GRB onset. 
Limiting the light curve analysis from the last significant WFC data point to the 
 second WFC upper limit, the flux decay is consistent with a power--law with slope 
$>1.9$, which however  is not consistent with the late afterglow data points. 
This would imply the presence of a bump in the time interval from 300~s to 
$\sim 3\times10^4$~s after the GRB onset.

%
%

\begin{figure}
\centerline{\includegraphics[width=8.5cm]{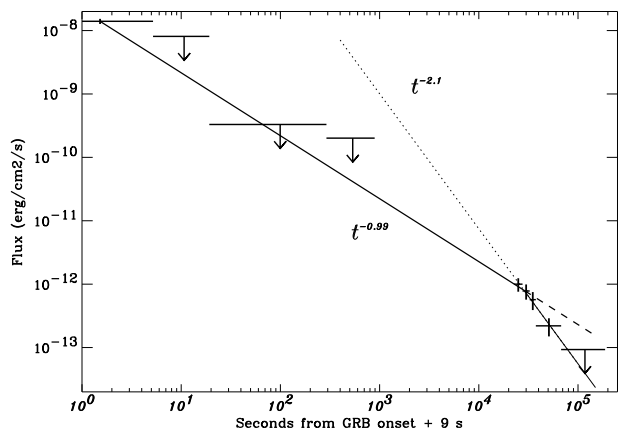}}
\caption[]{ 2--10 keV flux decay, assuming as origin the GRB onset
time plus 9~s.
The dashed line indicates the best-fit power--law, obtained by
including significant data points from both prompt and afterglow emissions,
but the last one. The solid line refers to the best-fit broken power--law.
}
\label{f:aft_lc1}
\end{figure}

Alternatively, taking as time origin that corresponding to 9~s after the GRB onset, we 
obtain the decay curve shown in Fig.~\ref{f:aft_lc1}.
This time origin corresponds to $\sim$ 50\% of the entire time duration of
the GRB, where the afterglow is expected to start (\cite{Frontera00}).

If we fit all the significant data points with a single
power--law (best fit index $\delta = 1.2$), the WFC upper limits 
are fulfilled, but the NFI upper limit is not consistent with this best fit curve. 
This suggests that a  break  somewhere in the light curve
should have occurred. 
If $\delta_1$ and $\delta_2$ are the power--law indices before and after the break time 
$t_b$, respectively, an estimate of $\delta_2$ is  $2.1^{+1.0}_{-0.6}$; for $\delta_1$
we fit the same points but the last significant one, since the break time seems to
occur somewhat earlier: it comes out $\delta_1=0.99_{-0.09}^{+0.04}$, where the
shallowest decay allowed ($0.90$) has been estimated by connecting the WFC point
with the WFC 3$\sigma$ upper limit from 30~s to 300~s; this solution fulfills all
WFC and NFI upper limits.
An estimate of the break time $t_b$ has been derived by intersecting the two
power--laws with their uncertainties. The result is $t_b=2.9^{+0.7}_{-2.4}\times10^4$~s,
(see Fig.~\ref{f:aft_lc1}).

Figure~\ref{f:sed} shows the Spectral Energy Distribution (SED) of the GRB afterglow.
In addition to the NFI measurements at different times, all the most sensitive
optical, IR and radio upper limits have been shown. The upper limits have been corrected 
for the Galactic absorption  using the color excess $E(B-V) = 0.024$ along the GRB 
direction (\cite{Schlegel98}), which implies an extinction $A_V = 0.07$ by using the
relationship $R_V = A_V/E(B-V) = 3.09 \pm 0.03$ by Rieke \& Lebofsky 
(1985)\nocite{Rieke85}. Following Cardelli et~al. (1989)\nocite{Cardelli89}
the extinction in the other optical colours is: $A_U = 0.13$, 
$A_B = 0.10$, $A_R = 0.06$, $A_I = 0.05$, $A_J = 0.02$,
$A_K = 0.01$. The corrected magnitudes have been converted into
flux densities $F_{\nu}$ ($\ergcms$ Hz$^{-1}$) according to
the calibration reported by Fukugita et~al. (1995)\nocite{Fukugita95}.

%
%

\section{Discussion}
\label{s:disc}
From the average spectral hardness, GRB010214 is a classical long GRB.
The spectral evolution of the prompt X--ray emission (hard--to--soft) is that
typical of GRBs (e.g., Frontera et~al. 2000)\nocite{Frontera00}, if, as above discussed,
we assume a photoelectrically absorbed {\sc pl} as the best model for the interval 
A (see Table~\ref{t:tab1}).
Also the power--law dependence of the GRB pulse width on energy above 7 keV is 
in agreement with similar investigations performed above 10 keV with the BATSE experiment 
(\cite{Fenimore95,Norris96}). The power--law index of the pulse width
behaviour
with energy ($\alpha_{\rm FWHM} = 0.28 \pm 0.13$) appears consistent with the range 
of values (0.37 to 0.46) found by Fenimore et~al. (1995)\nocite{Fenimore95} and Norris 
et~al. (1996)\nocite{Norris96} for a sample of bright BATSE bursts. It is however
significantly lower than the corresponding power--law indices  found by Feroci 
et~al. (2001)\nocite{Feroci01} for the X--ray richest GRB990704 ($\sim 0.45$) and by
Piro et~al. (1998a) \nocite{Piro98a} for GRB960720 ($\sim 0.46$). Our result confirms
the validity of the working scheme discussed by Norris et~al. (1996)\nocite{Norris96}
that pulses are the basic units in bursts.  A peculiarity however emerges below 7 
keV: the pulse width shortens while a secondary pulse rises; only the superposition
of both pulses gives a FWHM consistent with the extrapolation of the pulse width
behaviour at higher energies (see Fig.~\ref{f:FWHM_lag}). 

%
%
\begin{figure}
\centerline{\includegraphics[width=8.5cm]{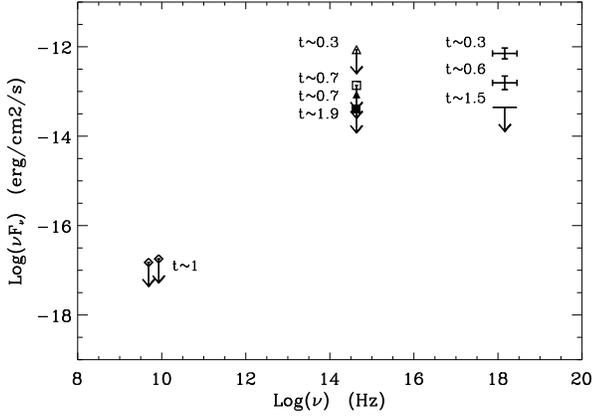}}
\caption[]{
Afterglow Spectral Energy Distribution: all available upper limits
are shown after correction for the Galactic extinction.
{\it open triangle}: Uemura et~al. (2001); {\it open square}: 
Klose et~al. (2001a)\nocite{Klose01a};
{\it filled triangle}: Hudec et~al. (2001)\nocite{Hudec01}; {\it filled square}:
Gorosabel et~al. (2001a)\nocite{Gorosabel01a}. 
The radio upper limits are reported by Berger \& Frail (2001)\nocite{Berger01}.
Times are expressed in days.
}
\label{f:sed}
\end{figure}

A much more peculiar feature of GRB010214 is the possible evidence of variable absorption 
of the promptly emitted X--ray radiation, from
$N_{\rm H} = 3.0^{+5.1}_{-2.0} \times 10^{23}$~cm$^{-2}$ during the first
6 s of the event down to a value consistent with the Galactic absorption 
($N_{\rm H}^{\rm G} = 2.66\times 10^{20}$~cm$^{-2}$) along the GRB
direction. So far, this feature has been observed only for GRB980329 (\cite{Frontera00}), 
GRB990705 (\cite{Amati00}) and GRB010222 (\cite{Zand01}).
As discussed by Lazzati \& Perna (2002),\nocite{LazzatiPerna02}
variable absorption might be the signature around the burst of an overdense cloud 
of radius $R<5$~pc and density $n>10^3$~cm$^{-3}$, similar to star formation globules within
molecular clouds (Bok globules). 
The true $N_{\rm H}$ at the GRB site could be determined only from the knowledge of the
GRB redshift $z$.
A tentative estimate of $z$ could be obtained as follows. Amati et~al. 
(2002),\nocite{Amati02} analyzing the
redshift-corrected average spectra of a sample of 12 \sax\ GRBs with known redshift,  
found a correlation between low-energy photon index $\Gamma_X$ and redshift $z$ 
and between isotropic gamma--ray energy $E_{iso}$ (evaluated in the 1--10$^4$ keV band)
and peak energy $E_p^{rest}$ in the GRB rest frame. Using the average value of
$\Gamma_X = 0.65^{+0.16}_{-0.19}$ (see Table~\ref{t:tab1}) and inverting the 
relationship found by Amati et~al. (2002) ($\Gamma_X = 2.464 (1+z)^{-0.78 \pm 0.18}$),
we get $z = 4.5^{+11}_{-2.3}$.
On the basis of this result, the minimum value of $N_{\rm H}$ at the GRB site during
the time interval A would be about $1\times10^{24}$~cm$^{-2}$.
However, the estimate of $z$ and thus of $N_{\rm H}$ are clearly tentative,
since the GRB sample used by Amati et~al. to derive their correlation, is still
small and should be confirmed by larger GRB samples.
The analysis of the afterglow light curve in the 2--10 keV energy band
(see Fig.~\ref{f:aft_lc0} and Fig.~\ref{f:aft_lc1}) has shown that either
the afterglow starts close to the origin of the main event with the power--law 
decay near its end, or the afterglow starts  after the peak time of the prompt 
gamma--ray emission. In the former case a bump in the afterglow light curve, something
like that found  by Piro et~al. (1998b) \nocite{Piro98b} for GRB970508,
is inferred; in the latter case
a break in the power--law decay should have occurred with an estimated break time of 
$2.9^{+0.7}_{-2.4}\times10^4$~s since the GRB onset.
Similar breaks in the light curve have been found for GRB990510 (\cite{Pian01}) and
GRB010222 (\cite{Zand01}). In these cases 
it could be established that the breaks are achromatic, given their contemporary
presence in the optical band.

Assuming a break, within the fireball model scenario, three different interpretations 
can be given: 
1) deceleration of the fireball into a wind environment (Chevalier \& Li, 1999, 2000);
\nocite{Chevalier99,Chevalier00}
2) transition from a relativistic to a Newtonian regime of the outflowing
matter (\cite{Dai99}); 3) deceleration of 
relativistic ejecta within a jet into a constant density medium (\cite{Sari99,Rhoads99}).
The mechanism 1) appears unlikely. Indeed,
according to the wind model by Chevalier \& Li (1999, 2000), assuming a density 
decrease $\rho(r) \propto r^{-2}$, where $r$ is the distance from the GRB site,
when the transition from fast to slow
cooling of the expanding fireball takes place, the expected break in the power--law
decay index at high frequencies is $\delta_2-\delta_1=1/4$, which does 
not match the observed break of $1.1^{+1.0}_{-0.6}$.
The mechanism 2) appears inconsistent with our data. Indeed,
if the steepening of the light curve is due to NRP, for a cooling frequency
$\nu_c < \nu_X$ after the break time, the expected power--law decay 
index is $\delta_2=(3p-4)/2$ and the photon index of the {\sc pl} spectral model is
$\Gamma = p/2 + 1$, where $p$ is the {\sc pl} index of the electron energy distribution.
From the estimate of $\delta_2$ we get $p=2.7^{+0.7}_{-0.4}$,
that would imply an expected value of photon index $\Gamma_{exp} = 2.4^{+0.3}_{-0.2}$
which is inconsistent with the measured value $\Gamma = 1.3^{+0.8}_{-0.6}$.
Also in the case $\nu_c>\nu_X$, this model is still incompatible with our data.
Indeed, the power--law decay indices before and after the
break time are $\delta_1=3(p-1)/4$ and $\delta_2=(15p-21)/10$, respectively: thus,
it is possible to estimate $p$ from the measure of $\Delta \delta = \delta_2 - \delta_1=
3/4p - 27/20$, and it comes out $p=3.3^{+1.3}_{-0.8}$. Replacing this value
in the expressions of $\Gamma=(p+1)/2$, $\delta_1$ and $\delta_2$, the following
estimates are obtained, respectively: $\Gamma_{\rm exp}=2.15^{+0.65}_{-0.4}$,
$\delta_{1{\rm exp}} = 1.7^{+1.0}_{-0.6}$ and $\delta_{2{\rm exp}} = 2.8^{+1.2}_{-0.9}$.
$\Gamma_{\rm exp}$ and $\delta_{2{\rm exp}}$ are only marginally compatible with the
corresponding measured values, while $\delta_{1{\rm exp}}$ is incompatible.

The model 3) accounts for the observed properties
of the X--ray afterglow, independently whether $\nu_c<\nu_X$ or $\nu_c>\nu_X$. 
Indeed, according to this model, after the break,
the expected $\delta_2 = p$ independently of the cooling frequency
position in the spectrum and $\Gamma = p/2 +1$ if $\nu_X>\nu_c$, while $\Gamma = p/2 +1/2$,
if $\nu_X<\nu_c$. From the estimated 
value of $\delta_2 = 2.1^{+1.0}_{-0.6}$, $\Gamma_{exp} = 2.0^{+0.5}_{-0.3}$ or 
$ = 1.5^{+0.5}_{-0.3}$, respectively, which are both consistent with the observed index 
$\Gamma = 1.3^{+0.8}_{-0.6}$. With the found value of $p = 2.1^{+1.0}_{-0.6}$, also 
the expected index of the decay law before the break $\delta_1$ 
($ = (3p-2)/4$ if $\nu_X>\nu_c$, otherwise $ = (3p-1)/4$) is consistent with the 
observations, independently whether  $\nu_c<\nu_X$ or $\nu_c>\nu_X$.
 
The assumption of a cooling frequency $\nu_c$ below the X-ray band is preferred
on the basis of the derived SED (see Fig.~\ref{f:sed}). Actually, if 
a high optical extinction is at the origin of the non--detection of the OT
(other possible interpretations of the GRB darkness will be considered below),
the true SED should rise the optical data points above the X--ray intensities. 

Alternatively, if a bump after an initial decay of the X-ray afterglow is assumed
(see Fig.~\ref{f:aft_lc0}), the contemporary  non--detection of an OT  is expected 
in the case in which the GRB occurs in a dense star--forming region and the afterglow
is dominated, minutes to day from the main event, by small--angle scattering off 
dust grains (\cite{Meszaros00}). According to this model, the initial afterglow light curve
would be that unscattered, while a steeper decay curve is expected after the bump. 
We observe a similar picture, but the initial light curve is much steeper ($\delta >1.9$)
than the mean steepness of the afterglow light curves ($<\delta> = 1.30\pm 0.02$)
(\cite{Frontera02}). In addition an IR counterpart,
expected to be detected from such dark events at several days distance from
the main event has not been reported.
Thus, even though this model cannot be ruled out, the break hypothesis and its
interpretation in terms of a spreading jet model appears more likely.
%
%
\begin{figure}
\centerline{\includegraphics[width=8.5cm]{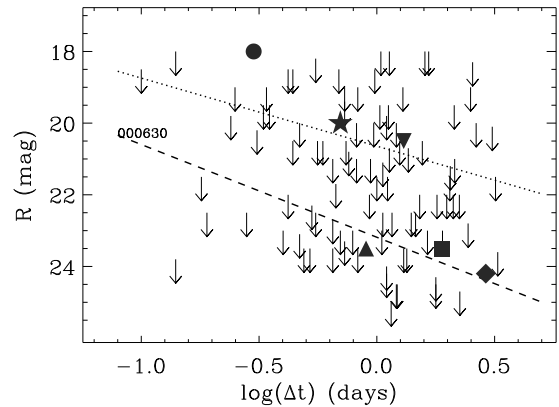}}
\caption[]{Limiting R-band magnitudes vs. observation
times for 55 GRBs without optical afterglow detection.
The dotted line shows the best linear fit obtained by
Lazzati et~al. (2002) between R magnitude and detection
time for a sample of detected optical afterglows.
The dashed line shows the power--law decay of the OT
of GRB000630 (\cite{Fynbo01}), one of the faintest OTs
within 1 day from the GRB. The filled symbols
refer to the following upper limits for GRB010214: 
the circle comes from Uemura et~al. (2001),\nocite{Uemura01}
the star from Klose et~al. (2001a),\nocite{Klose01a}
the triangle and the diamond from Rol et~al. (2001b),\nocite{Rol01b}
the upside down triangle from Hudec et~al. (2001),\nocite{Hudec01}
the square from Masetti et~al. (2001b).\nocite{Masetti01b}
Adapted from Fynbo et~al. (2001).}
\label{f:R-upp-lim}
\end{figure}

An open question is the density of the circumburst medium. At the onset of
the main event it is likely to be high as confirmed by the initial absorption of
the prompt X--ray emission. However the late prompt emission and afterglow spectra 
are consistent with a Galactic column density. Likely the GRB released energy
has almost completely ionized the initially absorbing cloud (\cite{LazzatiPerna02}). 
In this scheme however we would  expect to see the optical counterpart, as in the case 
of GRB980329 (\cite{Palazzi98}), which also showed an initial high
absorption (\cite{Frontera00}), and even a column density a 
factor 10 higher than the Galactic one during the afterglow phase (\cite{Zand98}). 
Thus a likely explanation of the GRB010214 darkness could reside in an intrinsic faintness
of the related OT and/or the high GRB redshift (centroid value of 4.5) estimated
by us. The non-detection of a radio counterpart of GRB010214 might also be related to an
intrinsic faintness of the source.

An important question in this respect is whether the non-detection of the 
OT is due to low sensitivity searches. To solve this
issue, we have compared the published optical upper limits of GRB010214 with
those of other dark GRBs and with the optical light curves
of faint OTs. In Fig.~\ref{f:R-upp-lim}
we show the result, adapted from Fig.~3 of the paper by Fynbo et~al. 
(2001).\nocite{Fynbo01}
Also shown is the best fit curve, obtained by Lazzati et~al. (2002)\nocite{Lazzati02}
for a sample of GRBs, which expresses the R magnitude as a function of the
detection time. 
The reported GRB010214 upper limits do not seem to suffer from inadequate
sensitivity observations when compared with other bursts. 

\section{Conclusions}
GRB010214 is a classical, long burst with detected X--ray afterglow, but
dark in the optical and radio band, despite sensitive and prompt searches.

The prompt emission shows the typical hard-to-soft spectral evolution
with a possible high column density
($N_{\rm H} = 3.0^{+5.1}_{-2.0} \times 10^{23}$~cm$^{-2}$) in excess of
the Galactic one ($N_{\rm H}^{\rm G} = 2.66\times 10^{20}$~cm$^{-2}$)
along its direction in the first 6 s.
Other three bursts were found to show similar variable absorptions:
GRB980329 (\cite{Frontera00}), GRB990705 (\cite{Amati00}) and GRB010222 (\cite{Zand01}).
Using the correlation between low-energy
photon index of the prompt emission and redshift found by Amati et
al. (2002), we have tentatively estimated the redshift $z\sim 4.5$ of GRB010214.

The 2--10 keV afterglow light curve suggests the presence
of a steepening or a bump  before the NFI observation with a break time
of $2.9^{+0.7}_{-2.4}\times10^4$~s from the GRB onset.
The  model which appears to better fit the data is
a spreading jet in either a typical ISM or in an overdense ionized cloud.
The latter environment is more likely, given the initially high $N_{\rm H}$.  
In the jet scenario, the power--law
index of the electron energy distribution has been derived
($p = 2.1^{+1.0}_{-0.6}$) and it is inside the range of values
found for a sample of \sax\ GRBs (\cite{Frontera00}).

\acknowledgements
This research is supported by the Italian Space Agency (ASI) and
Ministry of University and Scientific Research of Italy (COFIN funds).
We wish to thank the Mission Director L. Salotti and the teams of the \sax\
Operation Control Center, Science Operation Center and Scientific Data
Center for their efficient and enthusiastic support to the GRB alert program.

\end{document}